\documentclass[aps,twocolumn,showpacs,preprintnumbers]{revtex4}

\usepackage{graphicx,epsfig}
\usepackage{dcolumn}
\usepackage{bm}
\usepackage{hyperref}

\newcommand{\beq}{\begin{equation}}
\newcommand{\eeq}{\end{equation}}
\newcommand{\beqa}{\begin{eqnarray}}
\newcommand{\eeqa}{\end{eqnarray}}
\newcommand{\beqar}{\begin{eqnarray*}}
\newcommand{\eeqar}{\end{eqnarray*}}
\newcommand{\bra}[1]{\mbox{$\langle{#1}|$}}
\newcommand{\ket}[1]{\mbox{$|{#1}\rangle$}}
\newcommand{\diracsp}[2]{\mbox{$\langle{#1}|{#2}\rangle$}}

\def\I{{\rm i}}

\newcounter{saveeqn}
\newcommand{\alpheqn}{\setcounter{saveeqn}{\value{equation}}%
\stepcounter{saveeqn}\setcounter{equation}{0}%
\renewcommand{\theequation}{\mbox{\arabic{saveeqn}\alph{equation}}}}
\newcommand{\reseteqn}{\setcounter{equation}{\value{saveeqn}}%
\renewcommand{\theequation}{\arabic{equation}}}
\def\beql{\alpheqn \beqa}
\def\eeql{\eeqa \reseteqn}

\begin{document}

\title{Simplified and obvious expression of concurrence \\
in  Wootters' measure of entanglement of a pair of qubits}
\author{An Min WANG$^{1,2}$}

\altaffiliation{Supported by the National Natural Science Foundation of China under Grant No. 60173047, ``973" Project of China and the Natural Science Foundation of Anhui Province}

\affiliation{$^{1}$ Laboratory of Quantum Communication and Quantum Computing
and Institute for Theoretical Physics}
\affiliation{$^{2}$Department of Modern Physics,
University of Science and Technology of China \\
P.O. Box 4, Hefei 230027, People's Republic of China}



\begin{abstract}
 We obtain a simplified and obvious expression of  ``concurrence" in Wootters' measure of entanglement of a pair of qubits having no more than two non-zero eigenvalues in terms of concurrences of eigenstates and their simple combinations. It not only simplifies the calculation of Wootters' measure of entanglement,  but also reveals some its general and important features. Our conclusions are helpful to understand and use quantum entanglement further. 
\end{abstract}

\maketitle

Quantum entanglement is an essential and inherent  feature of quantum theory. Moreover, it is viewed as a useful resource in quantum information because it can be used in quantum teleportation \cite{QT}, quantum computation \cite{QC}, quantum cryptography \cite{QCG} and quantum error correction \cite{QEC}.  Therefore, it is interesting and important to quantify quantum entanglement. 

One of the best known measures of entanglement is entanglement of formation (EF) in a bipartite systems, which is proposed by Bennett {\it et.\ al.} \cite{Bennett}.  Its definition is
\beq
\label{EF}
E_{EF}(\rho_{AB})=\min_{\{p_i,\rho^i\}\in{\cal{D}}}\sum_{i} p_iS(\rho_B^i)
\eeq
where ${\cal{D}}$ is a set that includes all the possible decompositions of pure states $\rho=\sum_i p_i\rho^i$, $\rho_B^i={{\rm Tr}_A\rho^i}$ is the reduced density matrix of $\rho^i$ and  $S(\rho)$ is von Neumann entropy of $\rho$. 
For a pure state in bipartite systems, it is extensively accepted. However, in the case of mixed states, it is heavily dependent on the pure state decompositions and so far there is no a general algorithm to find the minimum one. In order to avoid this problem and improve EF's behavior in the mixed states, $via.$ Bennett {\it et.\ al.}'s suggestion and Wootters' development, so-called magic basis defined by four Bell's states $\ket{\Phi^{\pm}},\ket{\Psi^{\pm}}$ with particular phases 
\beql
\label{BellState}
\ket{e_1}=\ket{\Phi^{(+)}}=\frac{1}{\sqrt{2}}(\ket{00}+\ket{11}), \\
\ket{e_2}=\I\ket{\Phi^{(-)}}=\frac{\I}{\sqrt{2}}(\ket{00}-\ket{11}), \\
\ket{e_3}=\I\ket{\Psi^{(+)}}=\frac{\I}{\sqrt{2}}(\ket{01}+\ket{10}), \\
\ket{e_4}=\ket{\Psi^{(-)}}=\frac{1}{\sqrt{2}}(\ket{01}-\ket{10}),
\label{BellState4}
\eeql
plays a key role. One can rewrite EF for a pure state as a binary entropy function
\beq
\label{BEF}
H(z)=-z\log z-(1-z)\log(1-z)
\eeq
whose argument $z$ reads
\beq
z=\frac{1+\sqrt{1-C^2}}{2}
\eeq
where $C$ is called as ``concurrence". For a pure state,  it is equal to the norm of summation of square of coefficients of $\ket{\psi}=\sum_{i=1}^4\alpha_i\ket{e_i}$ in the magic basis, that is
\beq
\label{CinPS}
C=\left|\sum_{i=1}^4\alpha_i^2\right|
\eeq
Wootters and Hill's creative contribution is to find, for a mixed state $\rho$ of a pair of qubits, the concurrence can be defined by  \cite{Wootters} 
\beq
\label{CinMS}
C=\max(\lambda_1-\lambda_2-\lambda_3-\lambda_4,0)
\eeq
where the $\lambda_i$ ($\lambda_1$ is the maximum one) are the square roots of the eigenvalues of the following matrix
\beq
\rho^{1/2}\tilde{\rho}\rho^{1/2}=\rho^{1/2}\sigma_2^{A_1}\otimes\sigma_2^{A_2}{\rho}^*\sigma_2^{A_1}\otimes\sigma_2^{A_2}\rho^{1/2}
\eeq
and $\rho^*$ denotes the complex conjugation of density matrix $\rho$ in the computation basis $\{\ket{00},\ket{01},\ket{10},\ket{11}\}$. They proposed that the binary entropy function defined by eq.({\ref{BEF}) is also a measure of entanglement of mixed states (WE or BE) if its concurrence takes form (\ref{CinMS}), and proved that it is not larger than EF for the mixed state. This nice work is thought of the best result to quantify entanglement for bipartite systems consist of a pair of qubits in the known measures of entanglement so far. Obviously, it is a main task to calculate concurrence in order to obtain WE. Knowing concurrence is just knowing WE. Usually, one uses numerical calculation to obtain concurrence and then the values of WE. However, from the original form of concurrence or from its numerical result, it seems to be difficult to cognize and understand the general features of concurrence or  WE. Perhaps,  this will lead a limitation of WE's applications.  It is interesting to deduce out an analysis and obvious form of concurrence, as well as WE,  directly related with some physical quantities of the concerning quantum state. 

In our early works \cite{MyCPL}, we have found that a relation between the norm of polarized vector of reduced matrix  and  concurrence in the case of  pure states of bipartite systems. That is, 
\beq
\label{NPVandC}
\bm{\xi}^2=1-C^2
\eeq
In order to directly calculate concurrence of a pure state, we also prove that when the pure state is written as
\beq
\ket{\psi}=a\ket{00}+b\ket{01}+c\ket{10}+d\ket{11}
\eeq
its concurrence can be expressed by \cite{MyCPL}
\beq
\label{CforPS}
C=2|ad-bc|
\eeq
Moreover, for separable states, $\bm{\xi}^2=1$, and Bell's states  $\bm{\xi}^2=0$. Recently, we propose a sufficient and necessary separability criterion for pure states in multipartite and high dimensional systems \cite{MySC}, {\it i.e}, a pure state is (fully) separable iff the norm of coherent vector of  reduced density matrix of any $A_i$-partite arrives at the maximum value.  We are able to verify that for any cat state in multipartite systems,  the norm of coherent vector of  reduced density matrix of any $A_i$-partite takes the minimum value. It implies that so-called coherent vectors play an important role in the measures of entanglement. Therefore, we suggest a principle method to define and obtain the measures of entanglement in multipartite and high dimensional systems, at least in the case of pure states. In fact, according to our idea, we have tried to propose  the generalized entanglement of  polarized vector \cite{MyCPL}, the modified relative entropy of entanglement \cite{MyMRE} and the generalized entanglement of formation \cite{MyGEF} based on Bennett,  Wootters {\it et.al}'s as well as Vedral {\it et. al}'s ideas \cite{Bennett,Wootters,Vedral}. In this paper, we further would like to evaluate out how WE is dependent on the polarized vectors of reduced matrices or concurrences of eigenstates and their simple combinations, in order to analysis the general features of WE. Actually, it is also helpful to obtain an extended binary entropy of entanglement for multipartite and high dimensional systems in near future (in prepare).  

For simplicity, we only consider the case having no more than two non-zero eigenvalues in bipartite systems with a pair of qubits. Without loss of generality, set two eigenvectors of density matrix of a quantum state as
\beql
\ket{v_1}=a_1\ket{00}+b_1\ket{01}+c_1\ket{10}+d_1\ket{11}\\
\ket{v_2}=a_2\ket{00}+b_2\ket{01}+c_2\ket{10}+d_2\ket{11}
\eeql
which correspond to two non-zero eigenvalues, respectively denoting as $v_1$ and $v_2$,  the density matrix in eigen decomposition is just
\beq
\rho=v_1\ket{v_1}\bra{v_1}+v_2\ket{v_2}\bra{v_2}
\eeq
and then
\beq
\rho^{1/2}=v_1^{1/2}\ket{v_1}\bra{v_1}+v_2^{1/2}\ket{v_2}\bra{v_2}
\eeq
Through some calculations, we can obtain the eigenvalues of $\rho^{1/2}\tilde{\rho}\rho^{1/2}$ to be
\beq
\omega_\pm=x\pm\sqrt{y}
\eeq
where
\beqa
\label{xeq}
x&=&2 v_1^2|a_1 d_1-b_1 c_1|^2+ 2 v_2^2|a_2 d_2-b_2 c_2|^2\nonumber\\
& &+\frac{v_1 v_2}{4}\left|[(a_1+a_2)(d_1+d_2)-(b_1+b_2)(c_1+c_2)]\right.\nonumber\\
& &\left.-[(a_1-a_2)(d_1-d_2)-(b_1-b_2)(c_1-c_2)]\right|^2\nonumber\\
&=& \frac{1}{2}(v_1^2 C_1^2 +v_2^2C_2^2)+ \frac{v_1v_2}{4}|{\cal{C}}_+ -{\cal{C}}_-|^2\\
\label{yeq}
y&=&-\frac{v_1^2 v_2^2}{16}\left|\left\{[(a_1+a_2)(d_1+d_2)
-(b_1+b_2)(c_1+c_2)] \right.\right.\nonumber\\
& &\left.-[(a_1-a_2)(d_1-d_2)-(b_1-b_2)(c_1-c_2)]\right\}^2 \nonumber\\
& & -16(a_1d_1-b_1c_1)(a_2d_2-b_2c_2)\big|^2+x^2\nonumber\\
&=&x^2-\frac{v_1^2v_2^2}{16}\left|({\cal{C}}_+-{\cal{C}}_-)^2-4{\cal{C}}_1 {\cal{C}}_2\right|^2
\eeqa
Here, we have used the orthogonal and normalized relations of eigenvectors
\beq
\diracsp{v_i}{v_j}=\delta_{ij},\quad (i,j=1,2)
\eeq
and introduce the definitions of complex concurrences of pure states
\beqa
{\cal{C}}_i&=&2(a_i d_i-b_i c_i) \quad (i=1,2)\\
{\cal{C}}_\pm&=& (a_1\pm a_2)(d_1\pm d_2)-(b_1\pm b_2)(c_1\pm c_2)
\eeqa
In general, they are complex numbers. The usual concurrences of pure states are norms of our complex concurrences of pure states, that is, 
\beqa
C_i&=&2|a_i d_i-b_i c_i|,\quad (i=1,2)\\
C_\pm &=&|(a_1\pm a_2)(d_1\pm d_2)-(b_1\pm b_2)(c_1\pm c_2)|
\eeqa
where $C_\pm$ are concurrences with respect to the simple combinations of two eigenstates defined by
\beq
\rho_\pm=\frac{1}{2}\ket{v_1\pm v_2}\bra{v_1\pm v_2}
\eeq
From the definition of concurrence for a mixed state (\ref{CinMS}), it is easy to get
\beq
C^2=2x-2\sqrt{x^2-y}
\eeq
Substitute $x,y$ in eq. (\ref{xeq}) and eq.(\ref{yeq}), we have  the following theorem.

{\bf Theorem}\  For a quantum state of a pair of qubits having no more than two eigenvalues,  the square of concurrence can be expressed by
\beqa
C^2&=&(v_1^2C_1^2+v_2^2C_2^2)+v_1 v_2\frac{1}{2}|{\cal{C}}_+-{\cal{C}}_-|^2\nonumber\\
& &-v_1v_2\frac{1}{2}|({\cal{C}}_+-{\cal{C}}_-)^2- 4{\cal{C}}_1 {\cal{C}}_2|
\eeqa

This is our simplified and obvious expression of concurrence in a quantum state having no more than two non-zero eigenvalues. It means that the concurrence in such a mixed state is dependent on concurrences of eigenstates and their simple combinations. So is WE.  Obviously, for a pure state $v_1=1,v_2=0$ or $v_1=0,v_2=1$, our expression backs to  the definition (\ref{CinPS}). Note that we have had a relation (\ref{NPVandC}) between concurrence of a pure state and norm of polarized vector of reduced density matrix, as well as we know how to calculate easily  concurrence of a pure state by the coefficients of the state {\it via}. eq.(\ref{CforPS}), the square of concurrence of such a mixed state as well as WE can be written as an obvious function of polarized vectors of reduced density matrices of eigenstates and their simple combinations or the coefficients of two eigenvectors corresponding to two non-zero eigenvalues.

It must be emphasized that our theorem not only simplifies calculation of Wootters' measure of entanglement, but also reveals some its general and important features.  In the following corollaries and examples, we can more clearly see this advantage in our simplified and obvious expression. 

{\bf Corollary One} \  The square of concurrence has the upper bound and the lower bound expressed by
\beq
(v_1 C_1-v_2C_2)^2\leq C^2 \leq (v_1 C_1+v_2 C_2)^2
\eeq

Note that when $ 4|{\cal{C}}_1 {\cal{C}}_2|\geq |{\cal{C}}_+-{\cal{C}}_-|^2$, we have used the fact that 
\beqa
C^2 &\leq& (v_1 C_1-v_2 C_2)^2+v_1v_2({\cal{C}}_+-{\cal{C}}_-)^2 \nonumber\\
&=& (v_1 C_1+v_2 C_2)^2-v_1v_2 (4|{\cal{C}}_1 {\cal{C}}_2|-|{\cal{C}}_+-{\cal{C}}_-|^2) \nonumber\\
&\leq & (v_1 C_1+v_2 C_2)^2
\eeqa

{\bf Corollary Two} \  If eigenvectors only have real components (coefficients in the computation basis $\{\ket{00}$, $\ket{01}$,$ \ket{10}, \ket{11}\}$), then
when  $({\cal{C}}_+-{\cal{C}}_-)^2\geq 4{\cal{C}}_1 {\cal{C}}_2\geq 0$,  the square of concurrence arrives at the upper bound
\beq
C^2=(v_1 C_1+v_2 C_2)^2
\eeq
when $0\leq ({\cal{C}}_+-{\cal{C}}_-)^2\leq 4{\cal{C}}_1 {\cal{C}}_2$, the square of concurrence becomes
\beq
C^2=(v_1 C_1-v_2 C_2)^2+v_1v_2({\cal{C}}_+-{\cal{C}}_-)^2 
\eeq
while only ${\cal{C}}_1 {\cal{C}}_2\leq 0$, it reaches at the lower bound 
\beq
C^2=(v_1 C_1-v_2 C_2)^2
\eeq

{\bf Corollary Three} \  If ${\cal{C}}_+={\cal{C}}_-$, that is, $a_1d_2+a_2 d_1=b_1c_2+b_2 c_1$, the square of concurrence takes its lower bound
\beq
C^2=(v_1 C_1-v_2C_2)^2
\eeq

{\bf Corollary Four} \  If  $\rho_1=\ket{v_1}\bra{v_1}$ or $\rho_2=\ket{v_2}\bra{v_2}$ is separable, that is ${\cal{C}}_1=0$ or ${\cal{C}}_2=0$,  the square of concurrence is only dependent on another eigenvector and its probability, that is respectively
\beq
C^2=v_2^2 C_2^2\quad\mbox{or}\quad C^2=v_1^2 C_1^2
\eeq

It must emphasized that WE defined by a binary entropy function (\ref{BEF}) is a monotonic function of concurrence.  Thus, these corollaries, in special, corollary one,  tell us what is a possible minimum quantity of entanglement distilled from a mixed state and provide knowledge how to prepare a mixed state with a possible maximum quantity of entanglement. 

These corollaries give the simple forms of concurrence as well as WE  in many cases. It is just a very easy task to calculate them, in particular, for a lot of states including many interesting ones,  which obey the conditions in corollary two, corollary three and  corollary four. 

Corollary four points out an interesting problem that a lot of quantum states has the same entanglement based on Wootters' definition of binary entropy of entanglement. If it is real, then it will be helpful to keep entanglement against quantum error and distill the needed entanglement from various quantum state preparations. 
Corollary four also implies that when the probability related with entangled state is small enough and its concurrence is rather large, Wootters's measure of entanglement greatly decreases, comparing with Bennett {\it et. al}'s entanglement of formation, if  there is a precondition that at this time the eigen decomposition is just a minimum pure state decomposition. For example, $v_1=0.1, C_2=0$, we can compare WE and EF,  in the following figure:

\begin{figure}[h]
\includegraphics[width=2.8in]{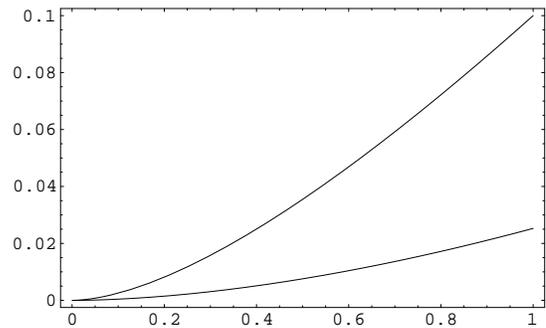}
\caption{The above curve is the entanglement of formation and the underside curve is Wootters' measure of  entanglement, which vary with $C_1$. }
\label{fig4}
\end{figure}

According to  Lewenstein and Sanpera's idea \cite{Lewenstein} to decomposition the concerning state into an entangled state and a separable state , above  precondition may be true. Therefore, the conclusion from corollary four is worthy of studying further. 

Now, let us discuss some useful examples.

{\it Example One} \  The mixture of two Bell's states. 

Because for all of four Bell's states $\ket{\Phi^{(\pm)}}$ and $\ket{\Psi^{(\pm)}}$ defined by eqs.(\ref{BellState}-\ref{BellState4}), it is easy to obtain that
\beql
{\cal{C}}(\Phi^{(+)})={\cal{C}}(\Psi^{(-)})=1\\
{\cal{C}}(\Phi^{(-)})={\cal{C}}(\Psi^{(+)})=-1
\eeql
we can easily evaluate out
\beq
\label{Cin2BM}
C^2=(1-2g)^2
\eeq
in  all of kinds of mixture of two Bell's states. Thus, their separability condition is just $g=1/2$. In the following, let us discuss them one by one. 

(1) $\rho=g\ket{\Phi^{(+)}}\bra{\Phi^{(+)}}+(1-g)\ket{\Phi^{(-)}}\bra{\Phi^{(-)}}$

Obviously, $a_+=1,b_+=c_+=d_+=0$ and $a_-=b_-=c_-=0, d_-=1$, that is ${\cal{C}}((\Phi^{(+)}\pm\Phi^{(-)})/\sqrt{2})=0$, we have eq.(\ref{Cin2BM}) in terms of corollary three. 

(2) $\rho=g\ket{\Phi^{(+)}}\bra{\Phi^{(+)}}+(1-g)\ket{\Psi^{(+)}}\bra{\Psi^{(+)}}$

Obviously, $a_+=b_+=c_+=d_+=1/2$ and $a_-=-b_-=-c_-= d_-=1/2$, that is ${\cal{C}}((\Phi^{(+)}\pm\Psi^{(+)})/\sqrt{2})=0$, we have eq.(\ref{Cin2BM}) in terms of corollary three. 

(3) $\rho=g\ket{\Phi^{(+)}}\bra{\Phi^{(+)}}+(1-g)\ket{\Psi^{(-)}}\bra{\Psi^{(-)}}$

Obviously, $a_+=b_+=-c_+=d_+=1/2$ and $a_-=-b_-=c_-= d_-=1/2$, that is ${\cal{C}}((\Phi^{(+)}\pm\Psi^{(-)})/\sqrt{2})=1$, we have eq.(\ref{Cin2BM}) in terms of corollary three. 

(4) $\rho=g\ket{\Phi^{(-)}}\bra{\Phi^{(-)}}+(1-g)\ket{\Psi^{(+)}}\bra{\Psi^{(+)}}$

Obviously, $a_+=b_+=c_+=-d_+=1/2$ and $a_-=-b_-=-c_-=- d_-=1/2$, that is ${\cal{C}}((\Phi^{(-)}\pm\Psi^{(+)})/\sqrt{2})=-1$, we have eq.(\ref{Cin2BM}) in terms of corollary three. 

(5) $\rho=g\ket{\Phi^{(-)}}\bra{\Phi^{(-)}}+(1-g)\ket{\Psi^{(-)}}\bra{\Psi^{(-)}}$

Obviously, $a_+=b_+=-c_+=-d_+=1/2$ and $a_-=-b_-=c_-=- d_-=1/2$, that is ${\cal{C}}((\Phi^{(-)}\pm\Psi^{(-)})/\sqrt{2})=0$, we have eq.(\ref{Cin2BM}) in terms of corollary three. 

(6) $\rho=g\ket{\Psi^{(+)}}\bra{\Psi^{(+)}}+(1-g)\ket{\Psi^{(-)}}\bra{\Psi^{(-)}}$

Obviously, $a_+=c_+=-d_+=0,b_+=1$ and $a_-=b_-= d_-0, c_-=1$, that is ${\cal{C}}((\Psi^{(+)}\pm\Psi^{(-)})/\sqrt{2})=0$, we have eq.(\ref{Cin2BM}) in terms of corollary three. 

{\it Example Two} \  Departure from Bell's states. 

We first consider the following four kinds of departure from Bell's state $\Psi^{(-)}$ to a mixture of it and a diagonal state, that is
\beq
B^D_i=p \ket{\Psi^{(-)}}\bra{\Psi^{(-)}}+(1-p) \ket{i}\bra{i}
\eeq
where $\ket{i}$ takes over diagonal states $\ket{00},\ket{01},\ket{10},\ket{11}$ respectively to $i=1,2,3,4$. It is easy to obtain that the square of their concurrences is   
\beq
\label{DBS}
C^2=p^2
\eeq

The simplest cases are that $i$ take 1 and 4. At this time, its two eigenvectors corresponding non-zero eigenvalues are $\ket{\Psi^{(-)}}$ and $\ket{i}, (i=1,4)$. Because $\ket{1}=\ket{00}$ and $\ket{4}=\ket{11}$ are both separable, we have the above equation in terms of corollary four. 

When $\ket{i}=\ket{2}=\ket{01}$, $B^D_2$'s two eigenvectors are
\beql
\ket{V_1}&=&\frac{1}{\sqrt{1+x_-^2}}\left(x_-\ket{01}-\ket{10}\right)\\
\ket{V_2}&=&\frac{1}{\sqrt{1+x_+^2}}\left(x_+\ket{01}-\ket{10}\right)\\
x_\pm&=&\frac{1-p\pm\sqrt{1-2p+2p^2}}{p}
\eeql
and their corresponding non-zero eigenvalues are
\beql
V_1=\left(1-\sqrt{1-2p+2p^2}\right)/2\\
V_2=\left(1+\sqrt{1-2p+2p^2}\right)/2
\eeql
Eigen decomposition of $B^D_2$ can be written as
\beq
B^D_2=V_1\ket{V_1}\bra{V_1}+V_2\ket{V_2}\bra{V_2}
\eeq
By means of our definitions, we can obtain
\beq
{\cal{C}}_1=\frac{2 x_-}{1+x_-^2},\quad {\cal{C}}_2=\frac{2 x_+}{1+x_+^2}
\eeq
Note that $x_-\leq 0$ and $x_+\geq 0$, we have ${\cal{C}}_1 {\cal{C}}_2\leq 0$. From our corollary two, it follows that
\beq
C^2= \left(V_1\frac{2 |x_-|}{1+x_-^2}-V_2\frac{2 |x_+|}{1+x_+^2}\right)^2
\eeq
Because $x_-x_+=-1$, we have $(1+x_-^2)(1+x_+^2)=(x_+-x_-)^2= 4(1-2p+2p^2)/p^2=4(V_2-V_1)^2/p^2$. Thus, it is easy to get 
\beq
C^2=p^2
\eeq
Likewise, when $\ket{i}$ take as $\ket{10}$, we have the same result. 

Then, we consider a departure from Bell's states to a mixture of it and its  orthogonal state, that is
\beq
B^D=q \ket{\Psi^{(-)}}\bra{\Psi^{(-)}}+(1-q) \ket{\chi}\bra{\chi}
\eeq
where 
\beq
\ket{\chi}= x_1\ket{00}+x_2\frac{1}{\sqrt{2}}(\ket{01}+\ket{10})+x_4\ket{11}
\eeq
Note that $|x_1|^2+|x_2|^2+|x_3|^2=1$. 
Thus, $B^D$'s two eigenvectors are just $\ket{\Psi^{(-)}}$ and $\ket{\chi}$ with eigenvalues $q$ and $1-q$ respectively. It is easy to obtain
\beq
{\cal{C}}_1=1,\quad {\cal{C}}_2=2(x_1x_4-x_2^2),\quad
{\cal{C}}_+-{\cal{C}}_-=0
\eeq
From corollary three, it follows that
\beq
C^2=[q-2(1-q)|x_1x_4-x_2^2|]^2
\eeq
Obviously, the condition that $B^D$'s is separable is 
\beq
|x_1x_4-x_2^2|=\frac{q}{2(1-q)}
\eeq
In particular, if $\ket{\chi}$ is separable, that is
\beq
x_1x_4=x_2^2
\eeq
we always have 
\beq
C^2=q^2
\eeq
It is the same as the result obtained by departure from Bell's state to a mixture of it and a diagonal state belonging to $\{\ket{00},\ket{01},\ket{10},\ket{11}\}$. We would like to know that why so many kinds of states have the same entanglement degree based on Wootters' definition of binary entropy of entanglement. In our previous paper \cite{MyMRE}, we found that there are differences between their entanglements for the departure from Bell's states to a mixture of it and a diagonal state based on our definition of modified relative entropy of entanglement.  It is interesting to research this problem further because we would like to know how to keep entanglement against quantum error and distill the needed entanglement from various quantum state preparations. 

In the end, we would like to point out that  our simplified and obvious  expression of concurrence has an obvious relation with the polarized vectors of reduced matrices or the concurrences of eigenstates as well as their simple combinations. Therefore, we conjecture that a similar relation in form might exist in the more general cases having more than two non-zero eigenvalues.  Its research is on progressing.

\end{document}